\documentclass[twocolumn,showpacs,preprintnumbers,amsmath,amssymb,superscriptaddress]{revtex4}

\usepackage{graphicx}
\usepackage{dcolumn}
\usepackage{bm}
\usepackage{epstopdf}
\usepackage{color}

\newcommand{\ef}{E_\mathrm{F}}

\newcommand{\vf}{v_\mathrm{f}}
\newcommand{\omegac}{\omega_\mathrm{c}}
\newcommand{\sigmaDC}{\sigma_\mathrm{DC}}

\newcommand{\omegaf}{\omega_\mathrm{F}}
\newcommand{\cg}{C_\mathrm{g}}

\newcommand{\cf}{c_\mathrm{f}}

\begin{document}

\preprint{APS/123-QED}

\title{Magnetoplasmonic Enhancement of Faraday Rotation in Patterned Graphene Metasurfaces}

\author{Michele Tamagnone}
\email{mtamagnone@seas.harvard.edu}
\affiliation{Laboratory of Electromagnetics and Antennas, \'Ecole Polytechnique F\'ed\'erale de Lausanne, Lausanne, Switzerland}
\affiliation{Harvard John A. Paulson School of Engineering and Applied Sciences, Harvard University, Cambridge, Massachusetts 02138, USA,}

\author{Tetiana M. Slipchenko}
\affiliation{Instituto de Ciencia de Materiales de Aragon and Departamento de Fisica de la Materia Condensada, CSIC-Universidad de Zaragoza, Zaragoza E-50009, Spain}

\author{Clara Moldovan}
\affiliation{Nanoelectronic Devices Laboratory, \'Ecole polytechnique f\'ed\'erale de Lausanne, Lausanne, Switzerland}

\author{Peter Q. Liu}
\affiliation{Department of Electrical Engineering, The State University of New York at Buffalo, Buffalo New York 14260, USA}

\author{Alba Centeno}
\affiliation{Graphenea SA, 20018 Donostia-San Sebasti\'an, Spain}

\author{Hamed Hasani}
\affiliation{Laboratory of Electromagnetics and Antennas, \'Ecole Polytechnique F\'ed\'erale de Lausanne, Lausanne, Switzerland}

\author{Amaia Zurutuza}
\affiliation{Graphenea SA, 20018 Donostia-San Sebasti\'an, Spain}

\author{Adrian M. Ionescu}
\affiliation{Nanoelectronic Devices Laboratory, \'Ecole polytechnique f\'ed\'erale de Lausanne, Lausanne, Switzerland}

\author{Luis Martin-Moreno}
\affiliation{Instituto de Ciencia de Materiales de Aragon and Departamento de Fisica de la Materia Condensada, CSIC-Universidad de Zaragoza, Zaragoza E-50009, Spain}

\author{J\'er\^ome Faist}
\affiliation{Institute for Quantum Electronics, Department of Physics, ETH Zurich, Zurich CH-8093, Switzerland}

\author{Juan R. Mosig}
\affiliation{Laboratory of Electromagnetics and Antennas, \'Ecole Polytechnique F\'ed\'erale de Lausanne, Lausanne, Switzerland}

\author{Alexey B. Kuzmenko}
\affiliation{Department of Quantum Matter Physics, University of Geneva, CH-1211 Geneva 4, Switzerland}

\author{Jean-Marie Poumirol}
\email{Jean-Marie.Poumirol@unige.ch}
\affiliation{Department of Quantum Matter Physics, University of Geneva, CH-1211 Geneva 4, Switzerland}

\date{\today}

\begin{abstract}
Faraday rotation is a fundamental property present in all non-reciprocal optical elements. In the THz range, graphene displays strong Faraday rotation; unfortunately, it is limited to frequencies below the cyclotron resonance. Here we show experimentally that in specifically design metasurfaces, magneto-plasmons can be used to circumvent this limitation. We find excellent agreement between theory and experiment and provide new physical insights and predictions on these phenomena. Finally, we demonstrate strong tuneability in these metasurfaces using electric and magnetic field biasing.

\end{abstract}

\maketitle

Graphene is considered a very promising material for non-reciprocal magneto-optical applications at microwave, terahertz and infrared frequencies \cite{GusyninSharapovCarbotte2007,Hanson2008aniso,CrasseeFaraday,CrasseeOrlitaPotemskiEtAl2012,FallahiPerruisseau-Carrier2012a,SounasCaloz2012,sounas2012novel,AvourisMagneto2012,ShalabyPecciantiOzturkEtAl2013,ShimanoYumotoYooEtAl2013,SounasSkulasonNguyenEtAl2013,UbrigCrasseeLevalloisEtAl2013,MartinMoreno2013,LinWangGaoEtAl2014,WangWangPuEtAl2014,HadadDavoyanEnghetaEtAl2014,TamagnoneFallahiMosigEtAl2014,SkulasonSounasMahvashEtAl2015,Isolator}.
Two of the most common non-reciprocal devices are isolators and circulators, and both are realizable starting from a Faraday rotator
\cite{dionne2005circular,ShalabyPecciantiOzturkEtAl2013,SounasSkulasonNguyenEtAl2013,TamagnoneFallahiMosigEtAl2014}. Faraday rotation (FR) observed in uniform graphene typically exhibits a maximum at low frequency ($<1$ THz), and is barely present at higher frequencies, apparently precluding applications above 3 THz.\cite{CrasseeFaraday,CrasseeOrlitaPotemskiEtAl2012,UbrigCrasseeLevalloisEtAl2013}. It was experimentally found that the magneto-optical response in transmission is enhanced at the plasmonic resonance frequency, in structures such as graphene dots \cite{AvourisMagneto2012}, antidots \cite{Liu}, and ribbons \cite{Yan2013}. The effect of magneto-plasmonic resonance on the FR are currently experimentally unexplored even if it was numerically demonstrated that such plasmonic structures should also induce a blue-shifting of the Faraday rotation maximum \cite{TamagnoneFallahiMosigEtAl2014, Tymchenko:2013iu}. \\
In continuous graphene, the impedance of the continuous monolayer is given for the two opposite circular polarisations by  \cite{FallahiPerruisseau-Carrier2012a} : 
\begin{eqnarray}
\label{eqn:eigencond}
Z_{\pm}=\sigma_{\pm}^{-1}=\sigmaDC^{-1}[1+i\tau(\omega\pm\omegac)]
\end{eqnarray}
where $\omega$ is the photon frequency, $\tau$ the carriers' scattering time, $\sigmaDC=e^2\tau|\ef|(\pi\hbar^2)^{-1}$ is the low-temperature low-frequency limit of graphene's conductivity for no magnetic bias, and $\omegac=e\vf^2B|\ef|^{-1}$ is the semi-classical cyclotron frequency. In such a system the maximum FR always appear at energies bellow the cyclotron resonance. To go even further, when considering the highly doped (and/or low magnetic field regime) where $\tau^{-1}$ is dominant over $\omegac$, the Faraday rotation will peak at zero frequency and will not extend above a cutoff frequency given by $\omega=\tau^{-1}$. Explaining the experimentally observed strong reduction of FR above 3 THz  \cite{CrasseeFaraday,CrasseeOrlitaPotemskiEtAl2012,UbrigCrasseeLevalloisEtAl2013}.

\begin{figure}
	\includegraphics[width=7cm]{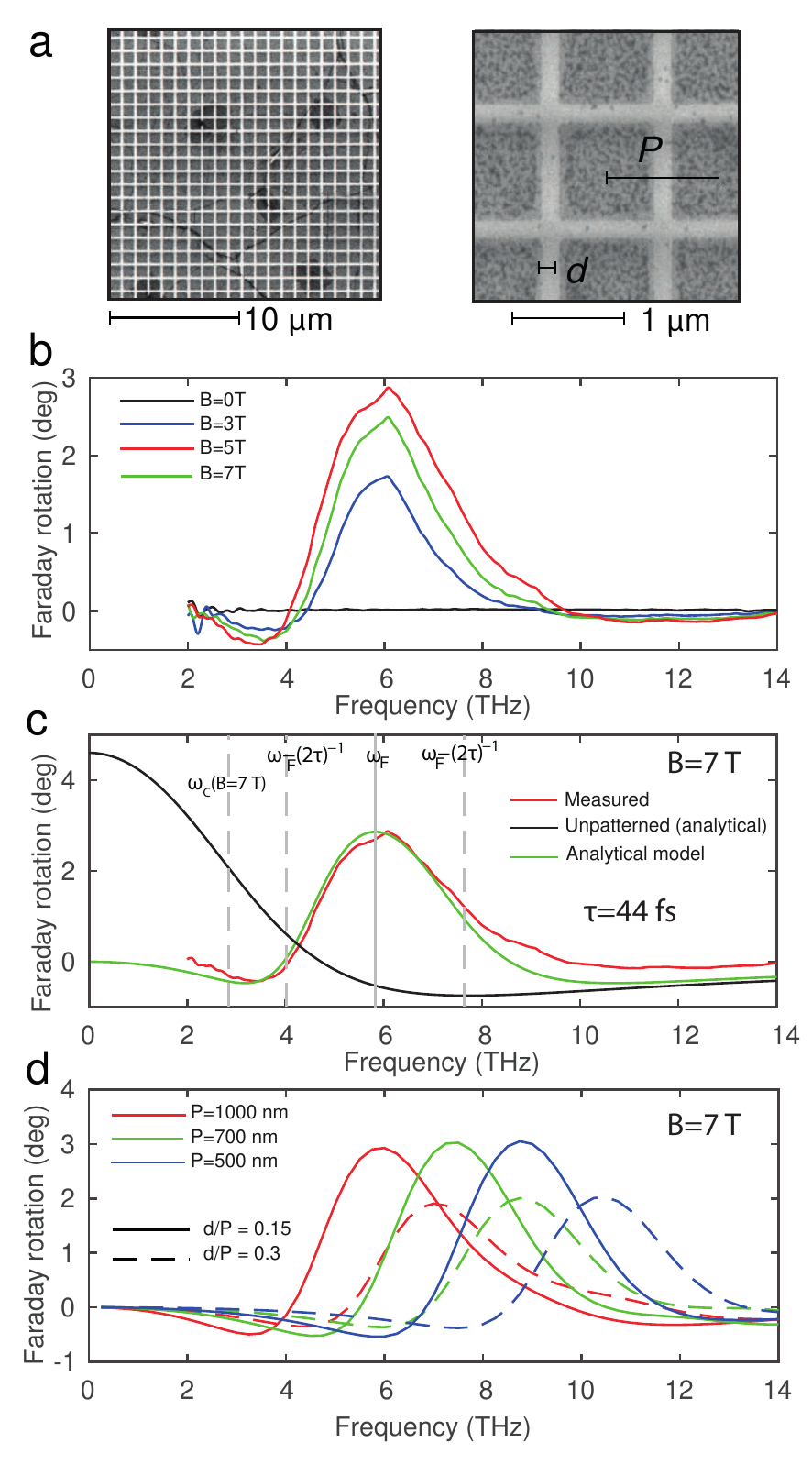}
	\caption{\label{fig:Figure1} (a) SEM pictures of the nano-patterned graphene square dot GSD lattice sample studied in this work. $P=1$ $\mu$m, d=150 nm. (b) Measured Faraday rotation on GSD sample at several magnetic fields up to 7 T. (c) Comparison between experimental data and the analytically predicted FR for $B=7$ T; expected broadening (central frequency) appears as dot (continuous) grey line. Expected $\omega_C$  in this system at B=7 T is shown as a grey dashed line. (d) Full wave simulations for various periods and fill ratios.  }
\end{figure}

Trying to circumvent these limitations, in this paper we have studied both experimentally and numerically the behaviour of the FR in three different patterned structures: a periodic array of graphene square dots (GSD), a graphene square anti-dot lattice (GSA), and a hybrid metal-graphene patterned structure (HMG). We have confirmed for all structures the presence of non-reciprocal magneto-plasmons blue shifting the Faraday rotation above the cyclotron resonance and compared the relative merits of the different geometries. 

All the samples measured in this paper are made from CVD graphene, transferred on oxidized high resistivity silicon wafers (oxide thickness: $t_\mathrm{ox}=280$ nm). The different pattern studied here were made using a combination of e-beam lithography, oxygen plasma etching and gold deposition. Faraday rotation spectra $\theta_F(\omega)$ are measured at room temperature ($T=280$ K), using linear polarized light following the procedure described in \cite{MOKKA}. Resolution is selected to remove the effect of the multiple reflection in the chip, equivalent to suppressing the phase coherence between multiple reflections. \\ 

A SEM image of the typical GSD is shown in Figure \ref{fig:Figure1} (a). The pattern consists of periodic squares with a periodicity $P=1$ $\mu$m and a distance between the dots $d= 150$ nm   \cite{FallahiPerruisseau-Carrier2012a}.

The FR measured in GSD is shown in Figure \ref{fig:Figure1} (b) for several magnetic fields up to 7 T. One can see that $\theta_F(\omega)$ in this structure displays a very different behaviour from the monotonous FR observed in continuous graphene and described in the introduction. For all non zero magnetic fields $\theta_F(\omega)$ exhibits a bell-shaped curve, centred around 6 THz. The amplitude of the FR increases with the magnetic field $B$, but neither the broadening nor the position of this maximum are affected by the variation of $B$.

To understand this behaviour we can generalize equation \ref{eqn:eigencond} for the patterned case. The bi-periodic pattern of etched gaps have the function of interrupting the path of surface currents on graphene adding a series capacitive term $\cg$, originating from the displacement currents. The impedance of the obtained graphene metasurface is then \cite{Whitbourn85}: 
\begin{eqnarray}
Z_{\mathrm{M}\pm} =\frac{\pi\hbar^2}{e^2|\ef|p} \left[\tau^{-1}+i\left(\omega\pm\omegac-\frac{e^2|\ef|}{\omega p\pi\hbar^2\cg}\right)\right]
\end{eqnarray}
where $p=\frac{8}{\pi^2}\left(1-\frac{d}{P}\right)^2<1$ is the pattern filling factor, reducing the overall conductivity tensor \cite{FallahiPerruisseau-Carrier2012a}. $p$ includes a current pattern factor $\cf=8/\pi^2$ which models the approximately sinusoidal current density profile on the patch, which gradually tends to zero at the edges \cite{balanis2005antenna}.
In analogy with continuous graphene a good estimation for the frequency of maximum Faraday rotation in the low mobility regime is  then found from the condition $\Im(Z_{\mathrm{M}+})+\Im(Z_{\mathrm{M}-})=0$, where $Z_{\mathrm{M}+}$ and $Z_{\mathrm{M}-}$ have the same modulus but the phase difference is maximized:
\begin{eqnarray}
\omegaf=\frac{e}{\hbar}\sqrt{\frac{\ef}{\pi p\cg}}
\end{eqnarray}
providing a non zero resonance value depending on the Fermi level and the chosen pattern but not on the magnetic field, as observed experimentally. $\omegaf$ can be understood as a plasmonic resonance arising between the kinetic inductance  $L_k=e^2|\ef|(\pi\hbar^2)^{-1}$ of graphene and the pattern-induced capacitance $\cg$. From the found impedance of patterned graphene, Faraday rotation is trivially found by analytically solving boundary conditions at the interface.  

Figure \ref{fig:Figure1} (c) compares the experimental data measured at $B=7$ T with the calculated FR  using an approximated analytical formula for $\cg$ of the considered pattern given in \cite{Whitbourn85}. A very good agreement is reached for the whole frequency range using $\ef=0.43$ eV and $\tau=44$ fs, thus capturing  the magnetic field independent resonance frequency $\omegaf$ and  bandwidth (from $\omegaf-(2\tau)^{-1}$ to $\omegaf+(2\tau)^{-1}$). To illustrate the shift of the FR above the cyclotron resonance, the black curve in Figure \ref{fig:Figure1} (c) is the calculated FR for continuous graphene using equation 1 and the extracted Fermi energy and scattering time. This allow us to see that the model also captures the amplitude of the FR and shows that, apart from the correction factor $p_{GSD}\approx 0.6$ which decreases the total Faraday rotation, patterned graphene retrieves the optimal Faraday rotation condition at the resonance frequency (found in uniform graphene for $\omega=0$), namely:
\begin{eqnarray}
Z_{\mathrm{M}\pm}(\omega=\omegaf)=p^{-1} Z_{\pm}(\omega=0)
\end{eqnarray}
Importantly, as shown in Figure \ref{fig:Figure1} (d), this tells us that, as long as the factor ratio $p$ is conserved, the amplitude of the Faraday rotation will be unchanged by a variation of the period $P$. Because $C_g$ is proportional to $P$ \cite{Whitbourn85}, the resonance frequency $\omegaf$ can be easily tuned over a large portion of the THz range without any loss of FR, and this independently of the cyclotron resonance energy.\\

\begin{figure}
	\includegraphics[width=7cm]{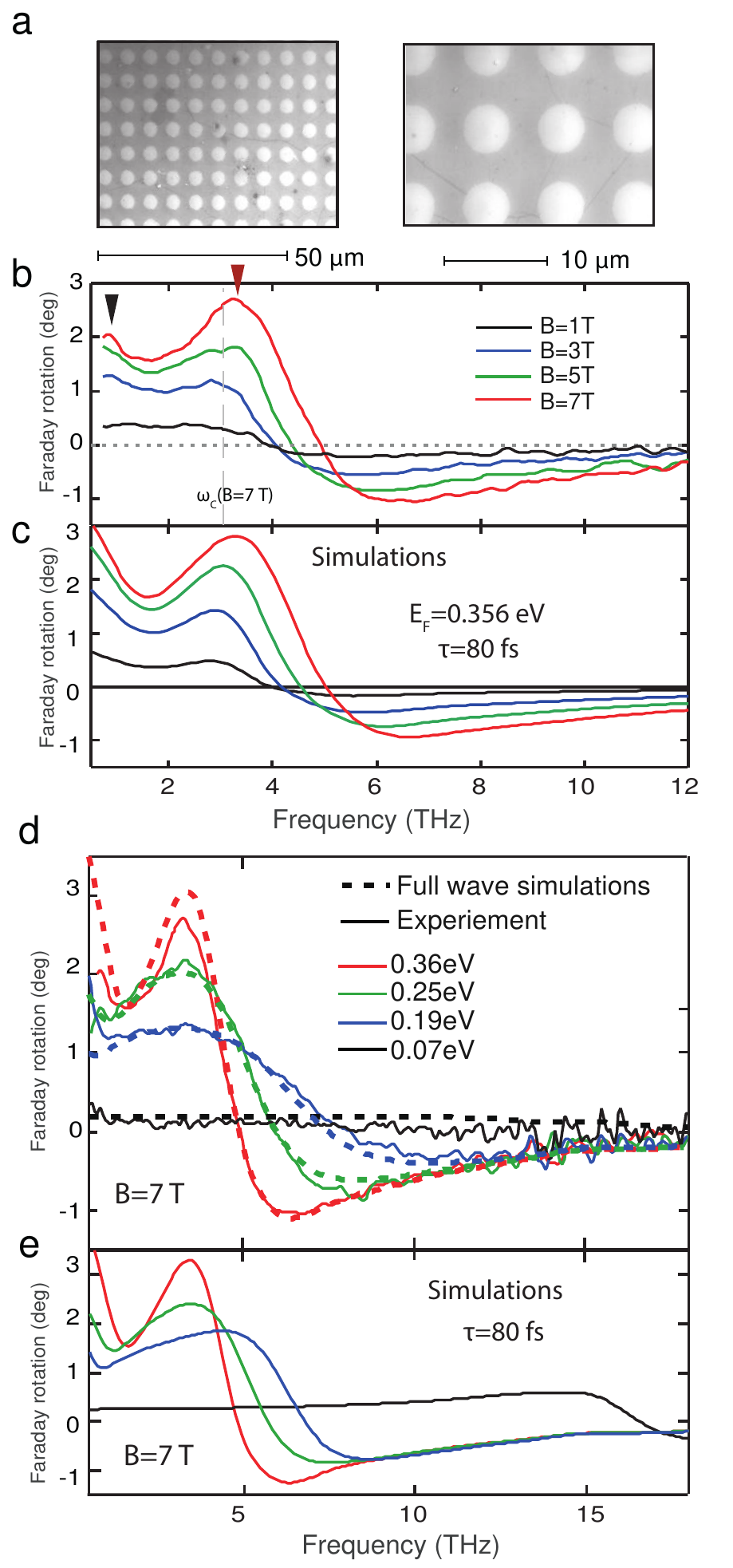}
	\caption{\label{fig:Figure2} (a) SEM pictures of the nano-patterned graphene square anti-dot lattice (GSA) sample studied in this work. $P=6$ $\mu$m, $D=4$ $\mu$m. (b) Measured Faraday rotation on GSA sample at several magnetic fields up to 7 T. Expected $\omega_C$  in this system at B=7 T is shown as a grey dashed line. (c) Simulated Faraday rotation for GSA with $\tau=80$ ps and $E_F=0.356$ eV. (d) Full lines: Faraday rotation measured at $B = 7$ T for different values of the $\ef$. Dashed line: Full wave simulations for the FR for fixed mobility $\mu=2500$ $\mathrm{cm}^2/(\mathrm{V}\cdot \mathrm{s})$ calculated for the corresponding value of $E_F$. (e) Simulated Faraday rotation measured at $B = 7$ T for fixed $\tau=80$ fs.}
\end{figure}

The main drawback of the GSD structures is that once fabricated, no $\textit{in-situ}$ tuning of the FR is possible. Firstly because the resonance frequency is magnetic field independent, and secondly because the Fermi level cannot be changed. To overcome these limitations, in the following we will focus on structures allowing more flexibility.  

Figure \ref{fig:Figure2} (a) is a SEM image of a typical GSA sample, with an anti-dot diameter $D=4$ $\mu$m and a periodicity $P=6$ $\mu$m. The apparent electrical continuity of this sample allows to apply a gate voltage.  

Figure \ref{fig:Figure2} (b) shows the Faraday rotation measured on GSA sample at gate voltage of $ V_G=136$ V (with respect to the charge neutrality point), corresponding to $\ef=0.356$ eV.  As the magnetic field increases, $\theta_F(\omega)$ displays two distinct maxima, the first one appears at low frequency (see black arrow) and seems to be centred at zero frequency, while the other one is centered around 3 THz (see red arrow). This behaviour can be understood by considering a superposition of the two behaviors described in the first part of this paper: (i) the GSA is an electrically continuous structure and as such Dirac carriers are free to move, thus the FR present a maximum at zero energy (similar to the one observed in continuous graphene); (ii) plasmonic resonances take place in this structure due to Bragg scattering on the periodic structure \cite{Liu}, and when it becomes coupled with the cyclotron resonance an enhancement of the FR is to be expected (and it is observed in the GSD structures). As $B$ increases, both maxima increase in amplitude reaching $2^{\circ} $ for peak A and nearly $3^{\circ} $ for peak B at $B=7$ T. It is interesting to note that with similar filling factors ($p_{GSA}=0.65$) both GSA and GSD present similar performances.

For a further understanding of this system, we performed finite-element electromagnetic simulations. Resulting numerical curves are shown in Figure \ref{fig:Figure2} (c), showing very good agreement with the experimental data. The shape, amplitude and frequency of the resonance are all well reflected  by the simulation, for all magnetic fields. From this fit we can extract an average scattering time of $\tau=80$ fs making the comparaison with the previous structure even more relevant. 

Let us now consider the $\textit{in-situ}$ tuning capabilities of this structure. Figure \ref{fig:Figure2} (d) shows the FR measured at 7 T for four different values of the Fermi level. As the Fermi level moves closer to the Dirac point the FR starts to decrease. $\theta_F(\omega)$ peaking at $2^{\circ} $ for $\ef=0.25$ eV and reaching a value close to zero for the whole experimental frequency range at  $\ef=0.07$ eV. Two phenomena take place simultaneously as $\ef$ decreases: (i) the density of carrier decreases, and the interaction between light and graphene becomes weaker, causing a smaller amplitude of the Faraday rotation (ii) the cyclotron resonance moves to a higher energy, rising as well the magneto-plasmon resonance. Figure \ref{fig:Figure2} (e) shows the simulated $\theta_F(\omega)$ for all the measured Fermi levels at a constant scattering time $\tau=80$ fs. The simulations clearly show that the value of the maximum FR shifts from 4 up to 15 THz as the Fermi level is decreased. It can be seen that the experimental data do not show the same behaviour, with no evidence of FR resonance for $\ef=0.07$ eV. Trying to reproduce the experimental data we realised that the scattering time what not a good fitting parameter, because as the Fermi energy decreases the effective mass of the graphene carriers changes $m=\ef/v_F^2$. The dash lines in Figure \ref{fig:Figure2} (d) show the simulated FR for all measured Fermi energies, taking into account a constant mobility $\mu=\tau/m=2500$ $\mathrm{cm}^2/(\mathrm{V}\cdot \mathrm{s})$ instead of a constant scattering time allowing a very good agreement with the experimental data.\\

\begin{figure}
	\includegraphics[width=6cm]{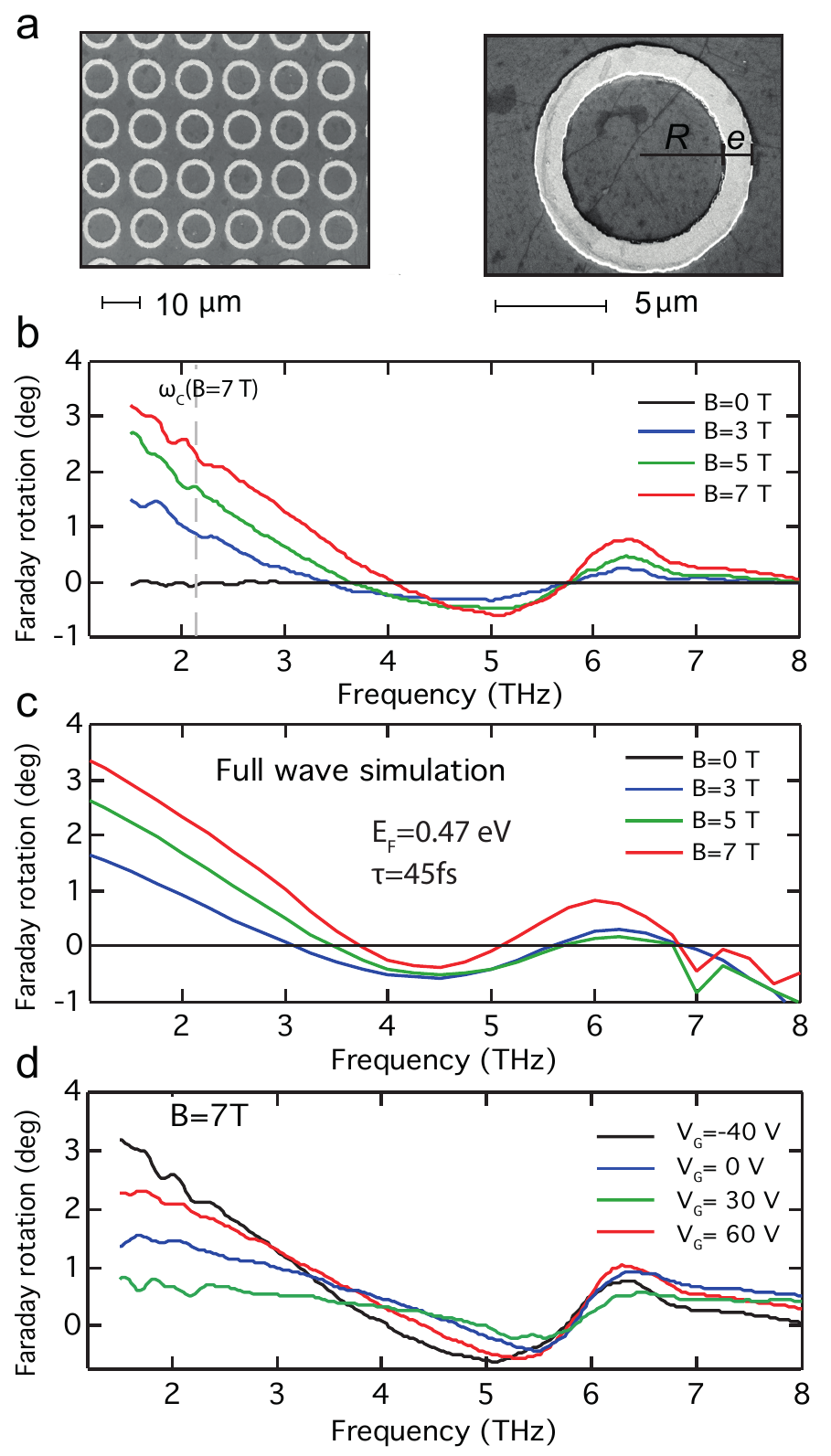}
	\caption{\label{fig:Figure4} (a) SEM pictures of the hybrid metal-graphene sample (HMG) sample $R=4$ $\mu$m, $e=1$ $\mu$m, $P=13$ $\mu$m. (b) Measured Faraday rotation at various fields, up to 7 T, for a fixed back gate voltage $V_g=50$ V. Expected $\omega_C$  in this system at B=7 T is shown as a grey dashed line. (c) Fit of the measurements with numerical full-wave simulations, for $E_F =0.47$ eV and $\tau=45$ fs. (d) Faraday rotation measured at $B=7$ T for different values of $V_G$.}
\end{figure}

Finally, the last structure studied in this paper is a periodic square lattice of gold rings covered by uniform monolayer graphene, as shown in Figure \ref{fig:Figure4} (a). The gold layer of 100 nm is evaporated directly on the silicon oxide and the graphene layer is transferred on it afterwards. The radius of the ring is $R=4$ $\mu$m, the ring width $e=1$ $\mu$m, and the pattern has a periodicity $P= 13$ $\mu$m.  An obvious  advantage of such HMG patterned structure is that no part of the surface graphene is removed. Consequently, the Fermi level can be tuned via gate voltage.

Figure \ref{fig:Figure4} (b) shows the results of the FR measurement for the HMG at a fixed back gate voltage $V_G=-40V$ for magnetic fields up to 7 T. Similarly to the previous system, two clear maxima appear in $\theta_F(\omega)$: the DC resonance is due to the free carriers and, the magneto-plasmonic resonance in this case lies at 6.3 THz. The value of the maximum Faraday rotation for peak B increases with increasing magnetic field and reaches almost $1^{\circ} $ at $B=7$ T.  

Results of the full-wave simulations are shown in figure \ref{fig:Figure4} (c). In this case the simulation is more complex due to the presence of the metal structures. To simplify the calculations, Maxwell's equations was not solved inside the gold, which was approximated as a thin film impedance. These simulations give us an insight on the physical principe taking place in the HGM: the presence of the gold rings in the vicinity of graphene affects the local electric field increasing it at the center of the ring. Graphene interacts with the enhanced field and the equivalent conductivity of graphene at this frequency becomes different for the left handed and right handed circular polarizations, causing Faraday rotation to appear. 

It can be seen that the model accounts well for the amplitude and frequency of both the experimentally observed modes. Interestingly, to reach such an agreement with the experimental data, the conductivity of the gold rings has to be decreased with respect to the nominal value. This is probably due to either discontinuities in the gold rings due to imperfect fabrication or to strong skin depth effects in metal at these frequencies. This may explain why, among the studied structures, the HGM structure presents the smallest enhancement of Faraday rotation of the studied structures. We believe that an improved fabrication process could result in better performances and would enable the structure to reach values of Faraday rotation similar to the two previous meta-structures. One can also see that the simulation predicts the presence of a third peak above 7 THz that appear clearly above 3 T, resonance that appear in the experimental data as a shoulder above the second resonance.

Figure \ref{fig:Figure4} (d) shows the FR measurement at constant magnetic field and for several gate voltages. Contrary to what has been observed in GSA, where the resonance frequency is tuned by the carrier density, the resonance frequency is fixed in HGM structures. The numerical model shows that the metallic pattern sets its own resonance frequency and dominates the response of the system. Hence, the resonance frequency does not appreciably depend on the charge carrier density in the graphene layer. Instead, the amplitude of the rotation is strongly affected by the carrier density, hence decoupling these two degrees of freedom.\\
 
In conclusion, we have demonstrated that by using plasmonic metasurfaces the Faraday rotation in graphene can be extended to the high terahertz band and potentially to the mid-IR, independently of the cyclotron resonance energy, thus paving the way to new non-reciprocal devices based on graphene. We compared the possibilities offered by three different types of structures. The periodic square arrays of graphene square dots (GSD), presents a resonant behavior acting in a specific frequency band. The frequency is ajustable over a large range of THz frequencies, in principle without any losses in the Faraday rotation. The periodic square array of graphene anti-dots (GSA) shows similar performances reaching $3^{\circ}$ of Faraday rotation, but on a broader frequency range due to multiple resonance frequencies. The amplitude and and frequency of these resonances can be tuned using gate voltage. Finally, the hybrid metal-graphene patterns structure (HMG) allows to decouple the frequency from the amplitude tuning.

This work has been financially supported by the Swiss National Science Foundation (SNSF) under grants 133583 and 168545, the Hasler Foundation under Project 11149 and the European Commission under Graphene Flagship (Contract No. CNECT-ICT-604391). We gratefully acknowledge discussions with Dr. Daniel Rodrigo Lopez.  We dedicate this work to the memory of Prof. Julien Perruisseau-Carrier, who passed away during the planning of the experiments.

\bibliography{biblio}

\end{document}